\documentclass[a4paper,12pt]{article}
\usepackage {graphicx}
\usepackage{epsfig}
\begin{document}

 \begin{center}
{\bf\Large Is it possible to obtain polarized positrons during
multiple Compton backscattering process? }
\end{center}

\begin{center}
{\large\bf A.P.~Potylitsyn}
\end{center}

\begin{center}
{\small\it Tomsk Polytechnic University \\
 Lenin Ave. 2a,  Tomsk, 634050, Russia \\
E-mail:pap@interact.phtd.tpu.edu.ru}
\end{center}

\bigskip

1. In existing projects of electron-positron colliders, the option
 of polarized electron and positron beams is considered [1,2].
While one can consider the problem of producing the polarized
electron beams with required characteristics as having been solved
[3], the existing approaches to polarized positrons generation
[4-7] do not provide required parameters. In quoted papers the
schemes were offered, in which by means of various methods a beam
of circularly-polarized (CP) photons with energy of $\sim $
10$^{1}$ MeV is generated to be subsequently used for producing
the longitudinally polarized positrons during the process of pair
creation in the amorphous converter.

In this paper an alternate approach is discussed - at the first
stage the unpolarized positrons are generated by the conventional
scheme (interaction of an electron beam with energy of $\sim
$10$^{1}$ GeV with an amorphous or crystalline converter), which
are accelerated up to energy $\sim $ 5 $\div$ 10 GeV and then
interact with intense CP laser radiation.

In the scheme of "laser cooling'' of an electron beam suggested in
the paper [8], electrons with energy of 5 GeV in head-on
collisions with laser photons lose their energy practically
without scattering. Thus, as a result of a multiple  Compton
scattering (MCS), the electron beam "is decelerated'' resulting in
some energy distribution, which variance is determined by the
electron energy and laser flash parameters. It is clear that the
laser cooling process will accompany also the interaction of
positrons with laser photons.

If we consider unpolarized positron beam as a sum of two fractions
of the identical intensity with opposite signes of 100\%
longitudinal polarization, its interaction with CP laser radiation
results in different Compton effect cross-sections for positrons
with opposite helicity. In other words, positrons polarized in
opposite directions lose a various part of the initial energy,
therefore, by means of momentum selection of the resulting beam,
it is possible to get a polarized positron beam with some
intensity loss.
\bigskip

2. Let us write the Compton effect cross-section of CP photons on
relativistic positrons after summing over scattered photon
polarization [9] (the system of units being used hereinafter is
$\hbar = m = c = 1$):
$$
 \frac{d\sigma}{d y} = \frac{\pi r^2_0}{x} \Big\{ \frac{1}{1-y}
+1- y - s^2 - \xi_{0z}\ P_c\  c y \ \frac{2-y}{1-y} \ -
$$
\begin{equation}
  - \ \xi_z P_c \Big[s_z  s c y + c_z \Big(\frac{y}{1-y} + y c^2
\Big)\Big] +
 \end{equation}
 $$
+ \xi_{0z}\xi_z \Big[s_z s(1+c^2 -y c^2)+  c_z c
\Big(\frac{1}{1-y} + (1-y)c^2\Big)\Big]\Big\} = $$
$$\frac{d\sigma_0}{d y} + P_c \ \xi_{0z} \
\frac{d\sigma_2}{d y} + P_c \ \xi_z  \frac{d\sigma_2}{dy} +
\xi_{0z} \ \xi_z \frac{d\sigma_3}{d y}\;.
$$

Here $P_{c}$  is the degree of circular polarization of laser
photons, $\xi_{0z}(\xi_z)$  is the spin projection of an initial
(final) positron on the  axis $z$ coincident with the direction of
the initial positron momentum, $r_{0}$ is the classical electron
radius. In (1) standard symbols are used [9]:
$$x = 2pk \approx 4\gamma_0 \omega_0\;,  \ y = 1-\frac{pk^{'}}{pk}
\approx \frac{\omega}{\gamma_0}\;,$$

 $\gamma_{0}$ is Lorentz
factor of an initial positron; $\omega_0(\omega)$  is energy of an
initial (scattered) photon. The factors $s, c$ are determined in
the known way [10]:
$$ s=2\sqrt{r(1-r)}\;, \ c=1-2r, \ r =
\displaystyle{\frac{y}{x(1-y)}}\;,
$$
where as factors $s_{z},c_{z}$ are obtained in going from the
coordinate  frame related to the momentum of positron scattered
through the angle $\theta_{e}$  and used in [9] to the initial
one:
$$ s_z = s - c \ \theta_e \;, \ \ \ c_z = c + s \ \theta_e\;.
$$

For an ultrarelativistic case $\theta_e = \frac{1}{\gamma_0} \
\frac{\sqrt{y(x - y - x y)}}{1-y} $, so with an accuracy of $\sim
\gamma_0^{-1}, \\ s_z = s, \ c_z = c $.

With the same accuracy, the cross-sections of spin-flip
transitions $d\sigma_{+ -}, \ d\sigma_{- +}$ from states with
opposite polarization signs ($\xi_{0z} = +1 \to \xi_z = -1 $ and
$\xi_{0z} = - 1 \to \xi_z = +1$) are equal. It means that the
Compton scattering process does not result in considerable
polarization of an unpolarized beam. It should be remarked that
the formula (1) is not the exact invariant expression (as well as
formula (12) in paper [9]). Both expressions may be written in the
invariant form with an accuracy of $\sim \gamma_0^{-1}$.  The
author's conclusion [11] concerning the possibility of
polarization of a positron beam as a whole  through MCS process
was  incorrect (it was based on the assumption that the magnitude
$ \int dy \Big[\frac{d\sigma_{+ -}}{d y} - \frac{d\sigma_{- +}}{d
y}\Big]$  presents an exact invariant which was calculated in the
rest frame of an initial positron,  see also [12]).

 3. As follows from (1), the total cross-section of positron
interaction with CP photons depends on spin projection
($\xi_{0z}$):
\begin{equation}
 \sigma = \frac{8}{3} \pi r_0^2 \  [(1-x) - P_c \ \xi_{0z}\ \frac{x}{4}]\;.
 \end{equation}

 In many cases of interest (laser cooling, for example) the
relation  $x\ll $ 1 is satisfied, therefore in (2) the terms $\sim
 x^{2}$ and higher are discarded. Let's write the cross-section
(2) for 100 \% right circular polarization of laser radiation (
$P_{c}$=+1) and for positrons polarized along the photon momentum
and in the opposite direction:
$$ \sigma_{\pm} = \int\limits_{0}^{x/1+x} \frac{d\sigma_{\pm}}{dy}\ dy
 \approx
 \frac{2\pi r^2_0}{x} \int\limits_{0}^{x/1+x} \Big[ 2-4\frac{y}{x}(1+y) +
 4\frac{y^2}{x^2}\big(1+2y\big)\pm \big(2y -4\frac{y^2}{x}\big) \Big]dy\;,
  $$
 \begin{equation}
 \sigma_+ = \sigma (P_c = +1, \ \xi_{0z} = +1)= \frac{8}{3} \pi r^2_0
 (1 - \frac{5}{4}x) =  \sigma_T \big(1-\frac{5}{4} x \big),
 \end{equation}
  $$ \sigma_{-} = \sigma (P_c = +1, \  \xi_{0z} = -1)=
     \sigma_T \big(1-\frac{3}{4}x\big)\;.
  $$

  Here $\sigma_T = \frac{8}{3}\pi r^2_0$
   is the classical Thomson cross-section. It is clear that due to
inequality of cross-sections (3), the positrons with various
helicities undergo the various number of collisions, that
eventually results in difference of average energies
$\bar{\gamma}_{\pm}$ of both fractions of the initial unpolarized
beam. With this distinction being sufficiently great, and the
variance of energy distribution for each fraction being enough
small, the polarized positron beam can be generated by means of
momentum selection.

4. In paper [13], in considering the MCS process by analogy with
passage of charged particles through a condensed medium, the
partial equations are derived that describe evolution of average
energy $\bar{\gamma}$ and energy straggling (distributions
variance) $\Delta$ for unpolarized electron beam passing through
an intense laser flash. The approximate analytical solution was
derived there as well:
$$
 \bar{\gamma} = \frac{\gamma_0}{1 +
  \displaystyle\frac{\sum^{(1)} l}{\gamma_0}}\;,
$$
\begin{equation} 
    \Delta = \frac{\sum^{(2)} l}{\Big(1+\displaystyle\frac{\sum^{(1)}
   l}{\gamma_0}\Big)^4}\;.
\end{equation}

In (4) $l$ is the laser flash length ("the thickness'' of light
target), $\sum^{(n)}$ is  the n-order moment of "macroscopic''
interaction cross-section:
\begin{equation}   
   \sum\nolimits^{(n)} = 2n_L \ \int\limits_0^{\omega_{max}}
   \omega^n \frac{d\sigma}{d\omega} \ d\omega = 2n_L \gamma_0^n
   \int\limits_{0}^{x/1+x} y^n \frac{d\sigma}{dy} \ dy\;.
   \end{equation}

 Here $n_{L}$ is the concentration of laser photons, that for
"short'' laser flash [8] is estimated as follows:
 \begin{equation}
     n_L = \frac{A}{\omega_0} \ \frac{1}{\pi
    r^2_{ph}l}\;,
    \end{equation}
 $A$ is the laser flash energy; $r_{ph}$ is  the minimum radius of the
laser beam.

Developing (1) as a series in powers of $x$ and retaining two
first summands, we get:
 $$\sum\nolimits^{(1)} = n_L \sigma_T \gamma_0 x \Big(1 - \frac{21}{10}x
    \Big)\;,
    $$
    \begin{equation}   
    \sum\nolimits^{(2)} = \frac{7}{10} n_L \sigma_T \gamma_0^{2} x^{2}
    \Big(1-\frac{22}{7} x \Big)\;.
    \end{equation}

After substitution of the found values for $\sum^{(n)}$ in (4) we
have:
\begin{equation}
\bar{\gamma} = \frac{\gamma_0}{1 + n_L \sigma_T l x \Big(1 -
{\frac{21}{10}} x\Big)}\;,
\end{equation}
\begin{equation}   
 \Delta = \gamma_0^2 \ \frac{{\frac{7}{10}} n_L \ \sigma_T \ l \ x^2
 \Big(1-{\frac{22}{7}} x \Big)}{\Big[ 1 + n_L \sigma_T l x
 \Big( 1 - {\frac{21}{10} x }\Big)  \Big ]^4}\;.
 \end{equation}

  Let's write the equation (9) in more evident form:
 \begin{equation}   
 \frac{\gamma_0}{\bar{\gamma}} = 1 + n_L \sigma_T  l x   \Big(1-
 {\frac{21}{10} x}\Big) = 1 + \frac{1}{2} k_0 \ x \Big(1 -
 {\frac{21}{10} x}\Big)\;.
 \end{equation}

 In approximation $x\ll $ 1 the quantity
$k_0 = 2n_L \ \sigma_T \ l $   corresponds to the  mean number of
scattered photons  per an electron of the initial beam (in other
words, the average number of collisions of an electron in passing
through the "light'' target). When expressing the photon
concentration $n _{L}$ in Gaussian laser beam in terms of Rayleigh
length $z _R$ and the photon wavelength $\lambda_{0}$, defining
minimum radius of the "light'' target
  $$ r_{ph}^2 = \frac{\lambda_0 \ Z_R}{2\pi}\;,
   $$
  one can readily see that the number of collisions is
independent of the laser wavelength directly:
$$ k_0 = \frac{16}{3} \
\alpha \ \frac{A}{mc^2}
   \frac{r_0}{Z_R}\;,
   $$
 here $\alpha$ is the fine structure constant.

The condition of the approximation applicability (4) (and,
therefore, (8) and (9) as well) is written as follows:
\begin{equation} 
   k_0 x^2 \ll 1 \;.
   \end{equation}

The second addend in brackets in (10) can be considered as a
correction related to the recoil effect of an initial electron.
This correction being neglected, from (10) one can derive the
classical result [14]:
 \begin{equation}
  \frac{\gamma_0}{\gamma_{f}} = 1 + \frac{l}{L_R}\;,
    \end{equation}
  \begin{equation}
  L_R = \Big( \frac{8}{3}\ r_0 \ \gamma_0
  \frac{4\pi^2}{\lambda_0^2} \
   a_0^2\Big)^{-1}\;.
   \end{equation}

In the relationship (12), $\gamma_{f} $ is the electron energy
after passing a laser flash;  the parameter $a_{0}^{2}$ is
determined by the formula (2) in the paper [14]:
  \begin{equation}
   a_0^2 = 3,66\cdot 10^{-19} \ \lambda_0^2(\mu) \ I (W/cm^2)\;.
   \end{equation}

Rewriting (14) in terms of previously derived quantities, we have
(for simplicity, here and in the formula (15) dimensional
quantities are used)
 \begin{equation} 
 a_0^2 = \frac{1}{\pi} \ \frac{r_0}{mc^3} \ I\lambda_0^2 =
 \frac{1}{\pi} \ \frac{r_0}{mc^3} \ n_L \ \omega_0 \ c \  \lambda_0^2 =
 2 \alpha \ \bar{\lambda}_e^2 \ \lambda_0 \ n_L\;,
\end{equation}
  $\bar{\lambda}_e$  is the Compton wavelength of an electron.

After substitution (14) in (12), the classical expression for
characteristic "thickness'' of the light target will be written in
terms of quantum characteristics of the laser flash
 \begin{equation}
\frac{1}{L_R} = \frac{32}{3} \ \pi r_0^2 \ \frac{\gamma_0
\omega_0}{m c^2} n_L = \sigma_T n_L x \;.
\end{equation}

 Thus, the relationship (10) in classical approximation coincides the formula
(12), if one compares the average electron energy
$\overline{\gamma}$ after the quantum process MCS with the final
energy of a particle continuously losing its energy by radiation
in travelling along a spiral trajectory in the field of plane ба
electromagnetic wave [15]. For electrons with initial energy
$E_{0}$ = 5 GeV having passed through a laser flash of following
parameters (see[8]): $\omega_{0}$ = 2,5 eV; $A$ = 5J; $r^2_{ph}$ =
4 $\mu m^2$, from (10) one can get ${\gamma_0}/{ \bar{\gamma}}
\approx $ 8.7.

Noteworthy is the reasonable agreement with estimates obtained by
V.~Telnov [8], though the criterion (11) is not satisfied in this
case.

Remaining terms proportional $x^2$ only the Eq.(9) may be written
as
$$\Delta/{\bar{\gamma}}^2 \simeq 7/10  \cdot
x{\bar{\gamma}} / \gamma_0 (1-{\bar{\gamma} / \gamma_0})$$
 which is rather close to Telnov's results [8].

5. As it was mentioned above, in ultrarelativistic case the difference
between probabilities of spin-flip transitions may be neglected, therefore,
in passing the unpolarized positrons through ба photon beam, the evolution
of each fraction of polarized positrons can be considered independently.

In this case,  the average energy of a fraction and variance may
be written in the full analogy with (4):
\begin{equation} 
\bar{\gamma}_{\pm} = \frac{\gamma_0}{1+
\displaystyle\frac{\sum_{\pm}^{(1)} l}{\gamma_0}}\;,
\end{equation}
\begin{equation}    
\Delta_{\pm} = \frac{\sum_{\pm}^{(2)}\ l}{\Big(1 +
\displaystyle\frac{\sum_{\pm}^{(1)} \ l}{\gamma_0}\Big)^4}\;.
\end{equation}

Here by $\sum_{\pm}^{(n)}$  the appropriate cross-section moments
are denoted:
$$ \sum\nolimits_{\pm}^{(n)} = 2 n_L \ \gamma_0^n \int\limits_0^{x/1+x} \
y^n \ \frac{d\sigma_{\pm}}{dy} \ dy\;.
$$
 The calculation of moments involved in (16) and (17) in
the same approximation  as before, gives the following result:
$$ \sum\nolimits_{+}^{(1)} = n_L \ \sigma_T \ \gamma_0 x \Big(1- \frac{8}{5} x \Big),
\ \ \sum\nolimits_{-}^{(1)} = n_L \ \sigma_T \ \gamma_0  x \Big(1
- \frac{13}{5} x \Big)\;;
$$
 $$ \sum\nolimits_{+}^{(2)} = \frac{7}{10} \ n_L \ \sigma_T \ \gamma_0^2 \
 x^2 \Big(1 - \frac{35}{14} x\Big), \ \  \sum\nolimits_{-}^{(2)} = \frac{7}{10} \
 n_L \ \sigma_T \ \gamma_0^2 \ x_0^2 \Big(1 - \frac{53}{14} x
 \Big)\;.
 $$
  Thus, the relative width of energy distribution in each fraction
is deduced from the relations:
\begin{equation} 
\frac{\sqrt{\Delta_{+}}}{\bar{\gamma_{+}}} =
\frac{\sqrt{\frac{7}{10} n_L   \sigma_T l x^2 \Big( 1-
\frac{35}{14} x \Big)}}{1 + n_L \sigma_T l x \Big(1  -
\frac{8}{5}\ x \Big) } = \frac{\sqrt{\frac{7}{20} \ k_0 \ x^2
\Big(1 -  \frac{35}{14} x \Big)}}{1 + \frac{1}{2} k_0 \ x \Big(1 -
\frac{8}{5} x \Big)} \;,
\end{equation}
 \begin{equation}    
 \frac{\sqrt{\Delta_{-}}}{\bar{\gamma_{-}}} =
 \frac{\sqrt{\frac{7}{10} n_L   \sigma_T l x^2 \Big( 1- \frac{53}{14}
x \Big)}}{1 + n_L \sigma_T l x \Big(1  - \frac{13}{5}\ x \Big) } =
\frac{\sqrt{\frac{7}{20} \ k_0 \ x^2 \Big(1 -  \frac{53}{14} x
\Big)}}{1 + \frac{1}{2} k_0 \ x \Big(1 - \frac{13}{5} x \Big)} \;.
\end{equation}

\begin{figure} [ht]
  \centering
\unitlength=1cm
\includegraphics[width=10cm,keepaspectratio]{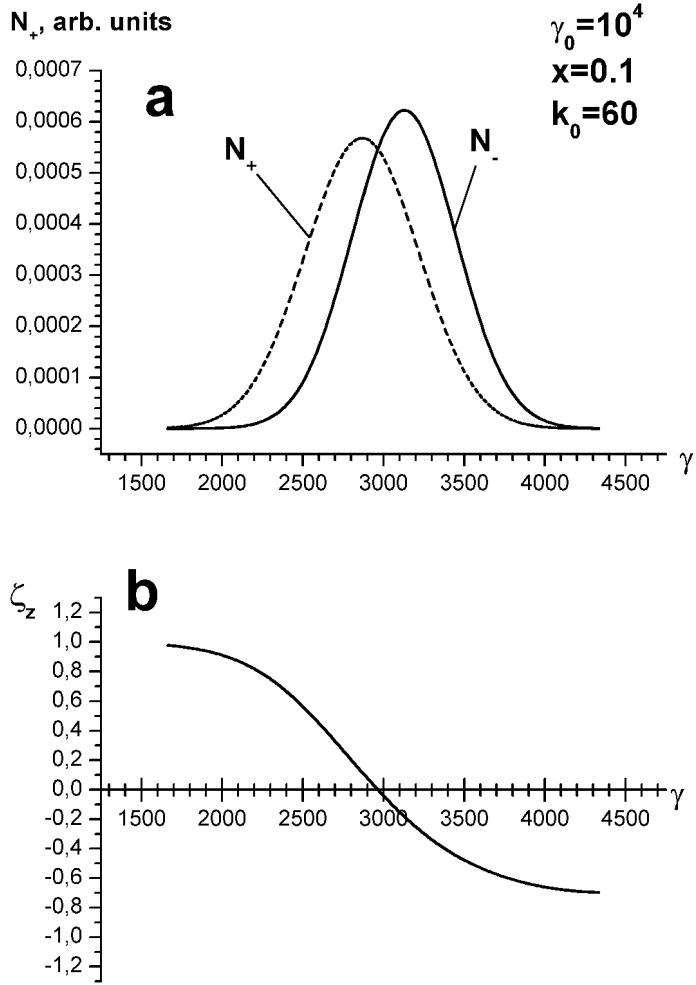}
 \parbox{15cm}{\caption{ a) Energy distribution of positrons polarized in
 opposite directions $N_{\pm}(\gamma)$ after passing a laser flash;
  b) the degree of longitudinal polarization $\xi_z(\gamma)$ versus positrons energy.
  }} \end{figure}

 \begin{figure} [ht]
  \centering
\unitlength=1cm
\includegraphics[width=8cm,keepaspectratio]{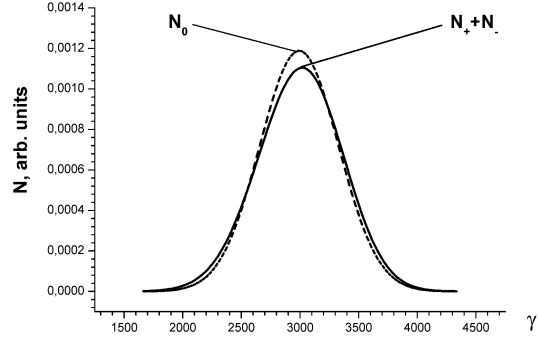}
\parbox{15cm}{ \caption{  Energy distribution of unpolarized particles
 $N_0 (\gamma)$ and the sum of distributions $N_+ (\gamma) + N_-(\gamma)$. }}
\end{figure}

\begin{figure} [ht]
 \centering
\unitlength=1cm
\includegraphics[width=9cm,keepaspectratio]{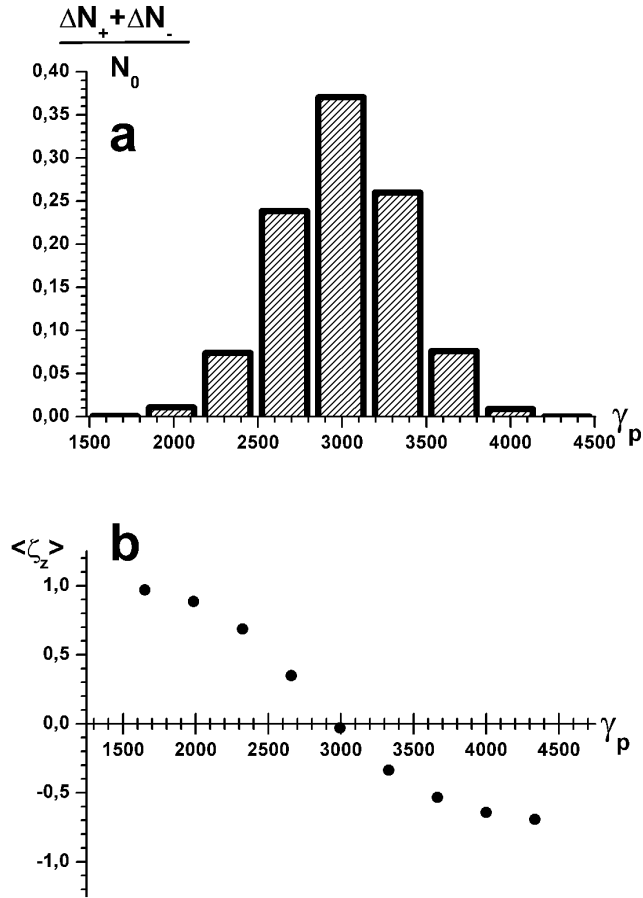}
\parbox{15cm}{\caption { a) Histogram of positron distribution
after momentum selection with acceptance  $\sigma = \sqrt{\Delta}$
(see Figure~2); \ b) the  degree of positron longitudinal
polarization after momentum selection.} }
\end{figure}

Figure~1a presents the distribution for each fraction after
passing the ба laser radiation with flash parameters: $A$ = 5J; \
$\lambda_0$ = 1 $\mu$m, \  $r_{ph}$= 4.2 $\mu m$($k_0$ = 60). The
distributions were approximated by Gaussians with parameters (17),
(18):
$$\bar{\gamma}_+ = 2868,  \
 {\sqrt{\Delta_+}}/{\bar{\gamma}_+} = 0.12\;;
 $$
  $$ \bar{\gamma}_- = 3129, \
  {\sqrt{\Delta_-}}/{\bar{\gamma}_-} = 0.10\;.
  $$

The degree of positron polarization being determined in the
ordinary way
  \begin{equation}
\xi_z (\gamma) = \frac{N_+ -  N_{-}}{N_+ + N_{-}}
\end{equation}

is shown in Figure~1b. Figure 2 presents the sum of distributions
of both fractions, which practically coincides the distribution
for the unpolarized beam with parameters (8), (9). It is evident
that only a small portion of positrons in the right (or left)
"tails'' of the sum of  distributions will have almost 100\%
longitudinal polarization. By means of momentum analysis with the
fixed acceptance $\Delta p/p = \Delta\gamma/\gamma $= const   in
proximity to a preset value $\gamma_p$ one can get a partially
polarized positron beam.

Figure~3 presents the polarization degree and intensity of the
positron beam resulting from the similar procedure, when after
passing a laser flash the beam had characteristics depicted in
Figure~1.
 For simplicity, the calculations were carried out for
uniform "capture'':
$$
  P =   \left\{ \begin{array}{lll}
  \rm const, &   \gamma_p - \displaystyle\frac{1}{2} \sqrt{\Delta} \leq \gamma
   \leq  \gamma_{p}  + \frac{1}{2}\sqrt{\Delta}
      \\
      &  & \\
    0 &  \mbox{ off}.
    \end{array}   \right.
    $$

   As follows from Figure~3, positrons with energy in the interval
$\gamma$ = 2660 $\pm$ 170 have average polarization $<\xi_{z}> \
\approx$ - 0.35, then in the interval $\gamma $ = 3330 $\pm $170,
\ $<\xi_{z}>~\approx $~0.34, with the positron intensity in each
"pocket"  reaching $\sim $ 24\% of the initial one.

It should be noted that in separating the final beam into two
parts  $(\gamma \leq \bar\gamma$  and $\gamma \geq \bar\gamma)$,
the intensity of both beams will be approximately identical (0,5
$N_0$ ), and  the average polarization decreases only slightly (
$< \xi_z >\ \approx \pm  $
  0.35).

6. As follows from Figure~2, the noticeable polarization can be
reached when the relation below is satisfied:
 \begin{equation} 
  \bar{\gamma}_{+} - \bar{\gamma}_{-} \sim \sqrt{\Delta_+}  + \sqrt{\Delta_{-}}
  \approx 2 \sqrt{\Delta}\;.
 \end{equation}

The last relation can be written in a simpler form for rather
"thick'' laser target (i.e. under condition $k_0 \gg$ 1). If in
addition to that, the inequality $x\ll $1 is satisfied, in this
case
 $$\bar{\gamma}_{+} - \gamma_{-} \approx \frac{2 \gamma_0}{k_0}\;.
$$
$$\frac{\sqrt{\Delta_+}}{\bar{\gamma_+}} \approx
 \frac{\sqrt{\Delta_-}}{\bar\gamma_-}
\approx  \frac{\sqrt{\Delta}}{\bar{\gamma}} \approx
\sqrt{\frac{7}{10} \ x \frac{1}{\frac{1}{2} \ k_0 \ x} } \sim
\frac{1}{\sqrt{k_0}} \;,
$$
  and the criterion (22) can be written in the form:
 \begin{equation} 
\sqrt{k_0} \ x \ \sim 1\;.
\end{equation}

In summary it should be noted that the results above were obtained
for the linear MCS process. For  an essentially nonlinear CS
process, when in each act of interaction a positron "absorbs'' $m
>$ 1 laser photons, the formulas (18)-(20) will remain valid only
in case of satisfying the inequality.
$$\bar{m} x \ll 1 \;. $$
Here $\bar{m}$ is the average of absorbed photons in one act of
interaction. Evidently that in this case moments
$\sum_{\pm}^{(n)}$  should be calculated in terms of nonlinear
Compton effect cross-section (see, for example, [16]). Comparing
spectra of scattered photons in linear and nonlinear processes
[17], one should expect that energy distribution variance of
positrons will be higher in the latter case, which can result in
increase of "overlapping'' the positron beam fractions $N_{+}$ and
$N_{-}$ in comparison with the linear case and, accordingly, in
decrease of positron polarization after the selection.

\end{document}